\documentclass{article}
\usepackage[utf8]{inputenc}
\usepackage{amsmath}
\usepackage{amssymb}
\usepackage{amsfonts}
\usepackage{mathtools}
\usepackage{amsthm}
\usepackage{xcolor}
\usepackage{url}
\usepackage{dsfont} %for indicator function
\usepackage{graphicx}
\usepackage{natbib}

\usepackage{tikz}
\usetikzlibrary{arrows.meta}

\usepackage[]{algorithm2e}

\usepackage{tabularx, environ}

\makeatletter

\newcolumntype{\expand}{}
\long\@namedef{NC@rewrite@\string\expand}{\expandafter\NC@find}

\NewEnviron{problem}[2][]{%
  \def\problem@arg{#1}%
  \def\problem@framed{framed}%
  \def\problem@lined{lined}%
  \def\problem@doublelined{doublelined}%
  \ifx\problem@arg\@empty%
    \def\problem@hline{}%
  \else%
    \ifx\problem@arg\problem@doublelined%
      \def\problem@hline{\hline\hline}%
    \else%
      \def\problem@hline{\hline}%
    \fi%
  \fi%
  \ifx\problem@arg\problem@framed%
    \def\problem@tablelayout{|>{\bfseries}lX|c}%
    \def\problem@title{\multicolumn{2}{|l|}{%
        \raisebox{-\fboxsep}{\textsc{\Large #2}}%
      }}%
  \else
    \def\problem@tablelayout{>{\bfseries}lXc}%
    \def\problem@title{\multicolumn{2}{l}{%
        \raisebox{-\fboxsep}{\textsc{\Large #2}}%
      }}%
  \fi%
  \bigskip\par\noindent%
  \begin{tabularx}{\textwidth}{\expand\problem@tablelayout}%
    \problem@hline%
    \problem@title\\[2\fboxsep]%
    \BODY\\\problem@hline%
  \end{tabularx}%
  \medskip\par%
}
\makeatother

\theoremstyle{plain}
\theoremstyle{definition}
\newcommand{\co}[2]{#2}

\title{Train Unit Shunting and Servicing: a Real-Life Application of Multi-Agent Path Finding}
\author{Jesse Mulderij\footnote{Delft University of Technology} \and Bob Huisman\footnote{NS, Utrecht, Netherlands} \and Denise T\"{o}nissen\footnote{Vrije Universiteit Amsterdam} \and Koos van der Linden\footnotemark[1] \and Mathijs de Weerdt\footnotemark[1]
}
\date{email: \texttt{j.mulderij@tudelft.nl}}
\begin{document}

\maketitle
\begin{abstract}
In between transportation services, trains are parked and maintained at shunting yards. The conflict-free routing of trains to and on these yards and the scheduling of service and maintenance tasks is known as the train unit shunting and service problem.
Efficient use of the capacity of these yards is becoming increasingly important, because of increasing numbers of trains without proportional extensions of the yards.
Efficiently scheduling maintenance activities is extremely challenging: currently only heuristics succeed in finding solutions to the integrated problem at all.
Bounds are needed to determine the quality of these heuristics, and also to support investment decisions on increasing the yard capacity.
For this, a complete algorithm for a possibly relaxed problem model is required.
We analyze the potential of extending the model for multi-agent path finding to be used for such a relaxation.
\end{abstract}

\section{Introduction}
    \co{TUSS informal introduction -- routing part of problem; importance of problem}
    The multi-agent path finding (MAPF) problem is an abstraction of routing problems such as that of robots or automated guided vehicles through an environment whilst preventing collisions.
    The state-of-the-art of complete solvers such as Conflict-Based Search~\citep{Sharon2015Conflict-basedPathfinding} and Branch-and-Cut-and-Price~\citep{Lam2019Branch-and-Cut-and-PricePathfinding} has recently progressed to facilitate routing as many as 50 agents on various benchmark instances~\citep{Lam2020NewFinding}.
    Therefore, this seems to be the opportune moment to make this abstract model more realistic, and start solving real problems.
    
    The application domain discussed in this paper is that of shunting and servicing trains, which considers all operations on trains outside the timetable.
    When at the end of the day trains have completed their service, they are parked on a nearby shunting yard, cleaned (sometimes also on the outside on a special washing track), repaired if necessary, and they receive regular checks and maintenance.
    After this, trains of the right type need to be ready at the right time to start executing their service in the timetable of the following day.
    For a small number of trains and a large shunting yard, finding feasible paths and schedules for the trains and their operations can be done by hand, but this becomes very difficult when a shunting yard is operating close to its capacity.
    
    \co{TUSS is not yet solved; algorithmically challenging; focus on feasibility} 
    
    \cite{VanDenBroek2016TrainApproach} has coined the name Train Unit Shunting and Service (TUSS) problem to indicate this application domain. It is an extension to the Train Unit Shunting Problem (TUSP), that originally focused on parking and matching arriving train units to departures only.
    TUSP has been investigated by several researchers in railway engineering and operations research, who provided mixed-integer programming (MIP) formulations for parts of the problem~\citep{Lentink2006ApplyingShunting, Kroon2008}, where typically the routing is considered separately. 
    Later several authors added one or more maintenance related service tasks to be scheduled simultaneously. 
    The full scope of the problem has been brought in by the ROADEF 2014 competition~\citep{Marcos2014}. This formulation is based on a real-life situation, and the results of that competition made clear that for practical application this problem could only be solved by heuristics~\citep{Geiger2018}. Since then these heuristics have been further developed by \cite{VanDenBroek2016TrainApproach}, \cite{Kremer2016ShuntingHeuristic}, and \cite{VanDeVen2019DeterminingSearch}.
    However, we cannot estimate the quality of the solutions found by these heuristics.

    %, unless a (likely relaxed) formulation can be optimally solvethat covers the dominant elements of the TUSS, while it can be considered as relaxation of it. The aim of this paper is to obtain such a complete formulation that includes the routing required for train shunting and servicing.
    %Initial comparisons of solutions found by the heuristic approaches and feasibility bounds obtained using exact (MIP) methods indicate that the routing --and avoidance of collisions-- is a very relevant part of the problem and should be integrated in these models. 
    
    \co{main question: how to use MAPF algorithms for TUSS, focus on optimal solving a relaxation}
    Only a complete algorithm can conclude that a given instance is infeasible and can provide a bound on the quality of heuristic methods.
    Moreover, such a complete algorithm is especially important for supporting decisions on investing in logistic capacity, since it can provide proof that capacity is inherently insufficient, and possibly even indicate the main bottlenecks.
    If such an algorithm is fast enough, it can also be used to support finding feasible solutions, for example by providing an optimistic heuristic in heuristic search, or by indicating combinations of choices as infeasible.
    Bounds on heuristics and advice on investment decisions can even be provided when abstract models are used, as long as such an abstraction is a relaxation and the algorithm is complete.
    
    The main objective of this work is therefore to identify whether MAPF models (and algorithms) can be used or extended to provide a relaxation of the TUSS problem.
    \co{our contributions: identifying differences and presenting these as step-wise extensions of MAPF, benchmark data?, relaxation is useful for practice -- no uncertainty}
    The contributions of this paper include an introduction of TUSS from the viewpoint of MAPF, and an analysis of similarities and differences to arrive at a relaxation of TUSS based on MAPF.
    
    \co{Setup of the paper}
    The next section contains a detailed description of the TUSS problem including the context and argumentation for the importance of providing bounds.
    Then, in Section~\ref{sec:MAPF} a model of MAPF is given and the similarities and differences from TUSS are analyzed, with in particular attention for aspects of the MAPF model that invalidate its use as a relaxation of TUSS.
    Section~\ref{sec:extensions} contains the consequential formal extensions of MAPF, including an analysis whether this addresses such a relaxation issue, or makes the model more precise, hopefully leading to better bounds.
    In a concluding section we reflect on these extensions and provide an outlook for next steps to be taken to let MAPF algorithms contribute to the practice of railway scheduling.

\section{Background}
First we provide more context and details on TUSS as it occurs in practice.
Then we highlight some relevant work from the literature.

%\subsection{Problem description}
% conceptual model
\co{ detailed description of TUSS practice (see introduction of thesis Roel, en anderen): doel: explain train shunting problem to laymen}

\subsection{Conceptual model}
This subsection describes the real-life problem, and those aspects that need to be incorporated in the formal models.

In a railway network the nodes represent major cities. For passenger transportation these hubs incorporate a station (or a few closely connected ones) and one or more yards. Many train services start and terminate at these hubs. One of the functions of a hub is to provide physical trains for these services, while terminating services cause a return flow. Then, at a yard trains are checked, cleaned, repaired, and parked. Functionally it resembles a repair shop with trains being the components to be delivered to the transportation process.

\begin{figure}[ht]
    \centering
    \includegraphics[width=1.0\textwidth]{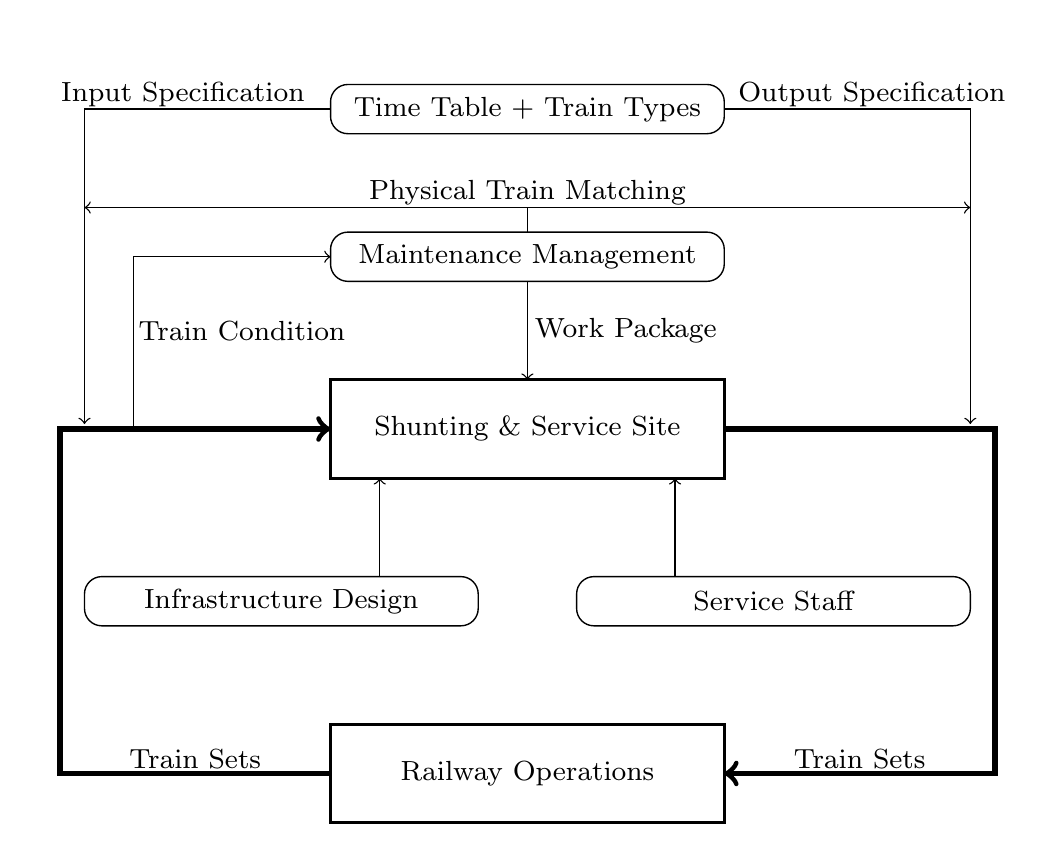}
    \caption{Interface between transport operation and a shunting \& service yard.}
    \label{fig:context}
\end{figure}

Figure \ref{fig:context} shows the interaction between the transport and yard processes. The timetable and related train-type assignment specifies the required output flow of the shunting and servicing site (i.e., the 'repair shop') and the return flow. 
It prescribes which types of trains, and in which composition, must be delivered at a specific station platform at what time. Vice versa it specifies the train units that become available when train services terminate. Although in some countries the timetable specifies the physical-train identifier, more often it specifies only the train type. Consequently the return flow contains different physical trains every day, even when a time table is executed repeatedly.

The maintenance program for a train type specifies the tasks to be executed with respect to the condition. For example, the need for a check depends on mileage. Because different trains arrive at a hub every day, and a train does not require daily the same maintenance services, the tasks to be scheduled vary over time.

At a yard certain tracks allow specific tasks to be executed. This resembles a flexible flow shop~\citep[e.g.][]{Yazdani2010}, where a physical train is represented by a job and the associated maintenance tasks to be executed by the operations. (Note that in the literature on flexible flow shops often an operation is what is called a task in this paper.) A set of tracks on which an operation can be performed is the equivalent of a set of similar machines. However, due to the rail infrastructure the routing is much more restricted than in a general flexible flow shop, and the intermediate stocks are limited due to track lengths.

Informally speaking, the objective of the \emph{Train Unit Shunting and Service Problem} is to find a feasible schedule for all activities required to take train units from a station platform to a yard, find conflict-free routes to tracks that are needed for the trains to get serviced and parked, and deliver the trains to the transportation process as specified, while respecting all kinds of resource constraints. In the following we briefly describe the sub-problems involved.

\subsection{Sub-problems to be addressed}
First, each train to be delivered needs to be mapped to one of the trains arriving and present on the yard. This is called the \emph{matching problem}. In general any train can be chosen for delivery if its type is compliant. However, sometimes a unique train is specified in order to direct it to a workshop.
Second, a \emph{resource-constrained scheduling problem} can be identified which represents the allocation of service resources. In addition to tasks that are known a priori, a schedule should include conditional tasks such as shunting between service tracks and staff that has to walk over the yard from one task to the other. Assigning a resource flow, that is the job sequence a resource unit follows, yields transportation tasks for trains and staff, which depend on spacial distances and site characteristics. 
Third, since staff needs to meet specific shift restrictions such as maximum number of subsequent hours driving a train, assigning a resource flow involves a special kind of \emph{staff scheduling and rostering problem}. 
Fourth, a \emph{parking problem} involves assigning a part of a track to each train for a certain time interval. Because a track can only host a relatively small number of trains, and each train type has its own length, this resembles a bin packing problem with additional constraints on the parking time (e.g., at platforms). 
Moreover, trains usually arrive and depart in some composition of two or more. To find feasible or optimal plans, \emph{decoupling and coupling} becomes relevant. 
Finally, a \emph{routing problem} is to be solved for trains to drive to and from the yard, while avoiding conflicts with regular passenger and cargo trains, and to visit the appropriate service and parking tracks as scheduled.

A typical approach is to solve these problems sequentially: for example, first do the matching for decoupled train units, then schedule the tasks and resources, allocate the parking places, find out where to couple train units, define the routing and finally assign personnel. This sequential approach has reached it limits, and is unable to solve today's hub problems due to the high traffic intensities. In particular, such sequential approaches cannot provide bounds. 
%Therefore integrated models have been developed. 

\subsection{Bounds for the logistic hub capacity}
Traditionally, railways produce timetables that are used for several months or sometimes even longer. 
Repeating the same solution for TUSS for every day (or week) is possible only by including sufficient slack in time and space, because, as described above, servicing requirements depend on the condition of individual trains. 
%To make a plan that can be executed repeatedly, while the trains do not need the same servicing every day, it must include slack in time and space. 
However, public transportation volumes keep on growing and 
unfortunately the rail infrastructure cannot be extended accordingly. Not only are typically the most important hubs situated in urban areas with often little space left to extend shunting yards, such extensions of the infrastructure determine to a high degree the total cost of railways, which should be reduced to let the industry be competitive in the future. 
If the planning of servicing and shunting can be done dynamically, such that only those activities are scheduled that are required based on up-to-date information about arriving trains and their condition, then the infrastructure can be used more efficiently. 
Furthermore, it is expected that such dynamic planning will also contribute to make the overall railway network more robust, because it increases its ability to compensate for disruptions. However, such an approach raises a new problem: How can the logistic capacity of a hub under dynamic planning be assessed to support strategic and tactical planning?

NS, the national railway operating company in the Netherlands, has developed a method to simulate dynamic hub logistics with on-line scheduling using local search. 
Although this has improved logistic analysis, the gap between the hub capacity of solutions found this way, and the upper bound found by optimization of a relaxed problem formulation remains significant. 
The currently used relaxed problem formulation covers all sub-problems mentioned before, except for the routing. It is expected that with routing included the relaxation gap will be reduced. The relaxed problem formulation should be complete, meaning that no feasible solution to the original problem is discarded. This also means that when the relaxed problem is infeasible, the original problem is infeasible as well.
This is the main motivation for applying MAPF to TUSS.

\subsection{Related Work}
\co{TUSP}
%To the best of our knowledge \cite{DiStefano2004AProblem} were the first to consider one of the sub-problems that is required for the shunting of train units. 
%They focus on the parking of train units such that the operations in the next morning can start as smoothly as possible. 
To the best of our knowledge, \cite{Freling2005ShuntingStation} provide the first attempt to capture a train unit shunting problem (TUSP) in a mathematical model. They split the problem in two sub-problems: the track assignment problem and the matching problem. 
%In the track assignment problem arriving trains are assigned to tracks and in the matching problem the arriving train units are matched to departures and the track. 
They are able to solve both models, and use real-life instances of the NS. \cite{Lentink2006ApplyingShunting} extend this formulation by not only looking at the assignments of train units to tracks and departures, but by also including the routing of the trains through the shunting yard. 
\cite{Kroon2008} and \cite{Lentink2006AlgorithmicPlanning} provide an integrated approach for the matching and parking problem. They aimed to include all possible configurations of train units and shunting yards in their models. 
\cite{Haahr2017OptimizationProblem} provides a comparison benchmark of multiple solution approaches for the TUSP, based on instances of the Danish State Railway and the NS. 
Besides using solution approaches from the literature they also introduce three new methods: a constraint programming formulation, a column generation approach, and a randomized greedy heuristic. 

\co{TUSS}
TUSS extends the TUSP by including service tasks, such as cleaning, regular checks, and maintenance. \cite{Geiger2018} show that for practical applications the TUSS problem can currently only be solved by heuristics. \cite{VanDenBroek2016TrainApproach} provides a simulated annealing algorithm that finds solutions for the TUSS problem. 
Furthermore, \cite{VanDeVen2019DeterminingSearch} combine machine learning with local search with the goal to provide bounds of the logistic hub capacity.
However, solving a (large) number of instances does not provide a guarantee that always solutions for instances with similar properties can be found, nor can be this be used to indicate instances as infeasible.
For this, a complete solver is required.
%Finally, \cite{Kamenga2019TrainStations} attempted to model the TUSS as an exact model.  However, their model still has many issues that needs to be solved before it could potentially be used in practise. 

%\cite{Kremer2016ShuntingHeuristic} student verslag ook een heuristic - nog niet geheel duidelijk wat het toevoegd.
%Lentink2006AlgorithmicPlanning was also the basis of the planning system that is currently used by the Netherlands Railways. OPG uses a sequential approach were first the matching and parking problem are solved and after that a decision is made on the routing of train units. Den Hartog  improves this  by solving the matching, the parking and the routing of the train units in an integrated step. 
%\cite{Cornelsen2007TrackAssignment} show that the parking problem is equivalent to coloring a conflict graph. 

Methods that can solve TUSS problems at scale do not exist, but for multi-agent path finding (MAPF), complete solvers such as Conflict-Based Search (CBS)~\citep{Sharon2015Conflict-basedPathfinding} and Branch-and-Cut-and-Price (BCP)~\citep{Lam2019Branch-and-Cut-and-PricePathfinding} can deal with instances of relevant size (e.g., 50--100 agents).
In both solvers, paths for individual agents are computed independently.
When such paths imply a collision, CBS considers these collisions pair-wise and creates two branches, one for each agent receiving priority over the other.
BCP uses a master LP that can select among an increasing number of shortest paths for the agents, and uses the dual of these collisions to define the cost of using that specific node in a next iteration of path finding.

\co{MAPF with task allocation: in TUSS tasks are maintenance and cleaning and come with the train (so allocation is not part of the problem)} 

MAPF with task allocation is an extension of MAPF where tasks have to be allocated to the agents. 
\cite{Ma2016} consider the assignment of (single) target task locations to agents before finding the routes.
In the more general multi-agent pickup and delivery problem (MAPD) \citep{Liu2019TaskDelivery} multiple tasks are assigned to agents and collision-free paths are planned for them to execute their tasks.
In this recent work, first a task sequence for each agent is found by solving a special traveling salesman problem, and then paths are planned accordingly. This approach is complete if the problem is well-formed, i.e., if (a) the number of tasks is finite, (b) the parking location of each agent is different from all task/end points, and (c) there exists a path between any two task/end points which traverses no other task/end points. MAPD does not include the matching problem and the related individual constraints on arrival and departure times, nor  the staff scheduling and (de)coupling sub-problems, but includes most of the other elements.
In TUSS the tasks are maintenance and cleaning; although they come with the train, there is some form of allocation, because sometimes there are different locations where the tasks can be performed.

%Ignoring the routing is outside the scope of this paper, but includes a broad field of research, such as on a repair shop for aircrafts~\citep{Bajestani2013}. Here, aircrafts are assigned to flights and the schedule of repair tasks is aimed at maximizing the flight coverage.
%when a failed aircraft enters the repair shop while the previous repair schedule is still under execution, reschedule the repair activities; techniques such as mixed integer programming (MIP); constraint programming (CP); logic-based Benders decomposition (LBBD), and a dispatching heuristic

%More generally, see e.g.~\citep{Wang2005}, for a survey on flexible flow shop scheduling.

Although the summary of related work above is far from complete, in our literature search we have not encountered any method that can solve the TUSS problem optimally at scale, and relaxations of TUSS appear not to have been a topic of study on itself thus far.

\section{The Multi-Agent Path Finding Problem} 
\label{sec:MAPF}

\co{Intro to problem def}
%As can be seen in the previous literature survey, many variants of the MAPF problem exist. 
%Some variants require grid worlds while others are defined for general graphs for example. 
It is important to be clear about the variant of MAPF that we assume to be given before we can propose (new) extensions. 
In this section, we provide a formal definition of the Multi-Agent Path Finding problem that is analogue to the one for example used by \cite{Lam2019Branch-and-Cut-and-PricePathfinding}.
    
    \co{Explain problem in detail (long paragraph)} 
Let $G=(L,E)$ be a connected, undirected graph that consists of a set of nodes $L$ called ``locations'' and a set of edges $E\subset L\times L$. Let $V= L \times \mathbb{Z}_{+}$ be the \emph{time-extended vertex set}, where a vertex $v \in V$ corresponds to a combination $v = (l,t)$ of a location $l\in L$ and a time $t\in \mathbb{Z}_{+}$. We denote $\hat{E} \subset V\times V$ to be the \emph{time-extended edge set}, where an edge $e = (v_{1},v_{2}) = ((l_{1},t_{1}),(l_{2},t_{2})) \in \hat{E}$ if and only if $t_{2} = t_{1} + 1$ and either $(l_{1},l_{2})\in E$ or $l_{1}=l_{2}$. We define the \emph{reverse edge} $e'$ of $e$ to be $e' = ((l_{2},t_1),(l_{1},t_2))$. We say that the directed graph $\hat{G} = (V,\hat{E})$ is the \emph{space-time graph} corresponding to $G$. 

Let us be given a set $A$ of which the elements $a\in A$ are called ``agents''. For each agent we are given a start- and goal location $s_{a}, g_{a}\in L$. A \emph{path} $p$ of length $k \in \mathbb{Z}_{+}$ for agent $a$ is a vector of locations $(l_{0},l_{1},\cdots,l_{k-1})$ where $l_{0} = s_{a}$ and $l_{k-1} = g_{a}$ and $((l_t,t),(l_{t+1},t+1)) \in E_t$ for all $t\in\{0,1,\cdots,k-2\}$. For $t>k$, agent $a$ remains at its goal location. Let $c:\hat{E}\to \mathbb{R}_{+}$ be a cost function on the edges of $\hat{G}$. For now we let $c(e) = 1, \forall e\in \hat{E}$. The cost $c(p)$ of a path $p$ is the sum of the traversed edges' cost. A feasible solution to the MAPF problem is a set of paths, one for each agent such that each vertex $v \in V$ is visited at most once and each edge $e$ and its reverse $e'$ are visited at most once in total.

Common objectives are to minimize the sum of individual costs (SIC) of the paths that agents take, or to minimize the latest time an agent arrives at its goal location, i.e., the makespan. Here we opt for SIC minimization.
Formally, the PSPACE-complete decision problem corresponding to the first of these objectives is described below.

\begin{problem}[framed]{Multi-Agent Path Finding (decision problem)}\label{prob:MAPF}
\textbf{Input}: & A graph $G = (L,E)$, the time extended graph $\hat{G} = (V,\hat{E})$ of $G$, a set of agents $A$, a vector of length $|A|$ of starting locations $\mathbf{s}\subset L$, a vector of length $|A|$ of goal locations $\mathbf{g} \subset L$, a cost function $c:\hat{E}\to 1$ on the edges and an integer $k$.\\
\textbf{Question}: & Does a set of paths $p_a$ exist (one for each agent $a\in A$), such that no pair of two paths visits the same node of $\hat{G}$ (resulting in a node conflict) and no pair of two paths visits both an edge $e\in \hat{E}$ and its reverse $e'$ (resulting in an edge conflict) and such that the total cost of all the edges in all paths satisfies $\sum_{a\in A} c(p_a) \leq k$?
\end{problem}
The minimization variant of this problem is finding the minimum $k$ for which the question is still answered affirmatively. Unlike most MAPF definitions that assume a grid-world, we specifically use a graph representation as it allows for a more precise modeling of a shunting yard.

\co{Explain similarities with TUSS, why is it worth it to use MAPF models for TUSS? (collision free routing)}
The MAPF problem has similarities with the routing sub-problem of the TUSS problem, but it can also solve part of the job scheduling sub-problem. 
Consider the following translation of (part of) an instance of TUSS to an instance of MAPF. 
The trains in the shunting yard can be seen as agents in the MAPF problem. 
The shunting yard can be modeled as a graph where each track is represented by a number of nodes. A natural choice for this number is to take the length of the track and divide it by the length of the shortest train, and rounding up the result. This allows for at least as much space for trains to move around on the graph is would be the case on the real shunting yard, making sure the formulation is a relaxation of the application. For service tasks that need to be performed at certain tracks, the goal is to find paths to these tracks for each of the trains that are not colliding (either on tracks or in resource allocation) during which the service tasks are completed.
In order to use MAPF as a relaxation, we plan to formulate a series of MAPF modifications that are all relaxations of the real-life version of the TUSS problem, but become progressively more restrictive, closing the gap between solutions to TUSS and bounds produced by the MAPF modifications.
%for any of these modified problems all preceding ones are a relaxation of the one it, and 
It is important to observe that the original MAFP formulation as given above is not a relaxation of TUSS. 
Therefore, it is important to first provide, as an essential starting point, a variant of MAPF that is a relaxation of TUSS, and then further refine this variant.

\section{Extensions}
\label{sec:extensions}

\co{intro, indicate some of the differences and the corresponding extensions} 
To guarantee this being a relaxation, it is essential that the first variant includes some basic model of the matching problem.
In the MAPF problem each agent has its own unique goal location $\mathbf{g} \subset L$ in the graph, while in TUSS any physical train of the appropriate type can be assigned to a departure. 
A relaxed version of this problem is to assume that all train units are interchangeable, and that the problem thus defines that each goal needs obtained by (at least) one of the agents.
% The MAPF model provides a solid basis in terms of routing trains on a shunting yard. There are however essential differences between the TUSS and MAPF problems. In this section, we discuss several of these differences and propose extensions of the MAPF model, with particular attention to enabling the use of optimal solutions to the resulting model as bounds. 
% First, we introduce train types into the model, along with constraints on arrival and departure times. 
Second, the direction of movement of trains is taken into account. Third, we propose to include a set of service tasks for each agent, resembling cleaning, repair and maintenance check-ups. Finally we mention several further possible extensions to reduce the gap between an optimal solution for the MAPF extension and feasible solutions to TUSS.

\subsection{Matching as a sub-problem}
\co{matching is part of the problem -- similarity to colors/types of agents, addition of arrival and departure times}
%An essential extension to the MAPF framework to make it a relaxation is to include the matching problem. 
In practice there are multiple types of train units. At the end of the timetable trains, which are compositions of train units, enter the shunting yard while at the end of the maintenance period trains depart, possibly in different compositions, to enter service the following day. 
%The composition of such a  train consists of an ordering of train types. 
It does not matter which specific train units fill the slot in the departing train, as long as they are of the prescribed type.
This property reminds us of robots that can perform specific tasks, which is modeled by the multi-commodity flow problem~\citep{even1975complexity}. 
For the moment, we presume all trains are decoupled, but address the extension of matching train units of certain types.
Another difference we address with this extension is that of differing arrival and departure times of trains.
In the classical MAPF problem, all agents start their routes at the same moment while trains leave and re-enter the transporting service at different moments throughout the day. The start- and goal nodes of an agent are therefore not only location dependent, but also time specific. 
%Furthermore, departures only specify that a train unit from a certain type enters transportation service at a certain time and location, in practice it does not matter which specific train unit of that type departs the shunting yard. 
We assume that for each train type, as many train units arrive as depart.

\co{model}
Every agent $a$ is assigned a type $o \in O$. The expected time $t \in \mathbb{Z}_+$ and location $l\in L$ of an agent's arrival is accurately represented by the vertex $(l,t)\in \hat{G}$. The start locations of agents are thus replaced with starting vertices in the time-extended graph. The goal locations are also replaced by vertices such that they carry a time specification. Goal vertices are also no longer attached to a specific agent. Instead, a set of goal vertices $g=\{g_{1}=(l_1,t_1,o_i), g_{2},\hdots,g_{n}\}$ is introduced, where $o_i$ is the specified train unit type that departs at that vertex. An agent of type $o$ can only depart from a goal vertex of the corresponding type. In the end, all departures have to be satisfied. Thus, a matching problem between the agents and the departures has to be solved implicitly or explicitly.

\co{relaxation argument}
This extension is important, because without it agents have fixed goal locations. This would effectively limit the options and might exclude practically feasible solutions. The model would then no longer represent a relaxation of the TUSS problem.
Ignoring the coupling and decoupling, and thus treating train units as agents does not reduce the number of options.
The proposed extension is thus a relaxation of TUSS.
Even the case where all agents are considered to be of the same type, all arrivals are at time 0, and all goals at the latest time in the input is still a relaxation, and could be a good start point.

\subsection{Direction of movement}
\co{Introduce what is lacking in the mapf model, why is this extension important?}
A possible next step in refining the relaxation is keeping track of the direction of movement of an agent. This extension is added to make the model more precise, resulting in a better capacity bound. When a train moves in a given direction, it means that the train driver must be located on the front-end of the train. Therefore, a train cannot simply reverse direction. The train driver has to walk towards the other side of the train, this process takes some time.
Ignoring this may lead to infeasible solutions because of missed deadlines, but does not remove potentially feasible solutions from the modeled search space.

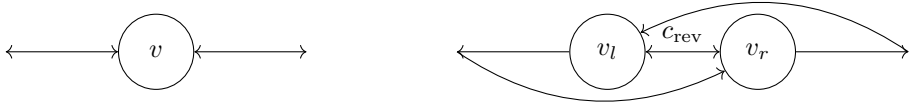
\begin{figure}
    \centering
\begin{tikzpicture}[scale=2,auto=center]%,every node/.style={circle, draw, minimum width = 1.8 cm}
    \node[style={circle,draw},minimum width = 1 cm] (v) at (0,0) {$v$};
    
    \path[<->] +(-1,0) edge node[above] {} (v);
    \path[<->] (v) edge node[above] {} +(1,0);
    
    \node[style={circle,draw},minimum width = 1 cm] (v_l) at (3,0) {$v_l$};
    \node[style={circle,draw},minimum width = 1 cm] (v_r) at (4,0) {$v_r$};\

    \path[->] (v_l) edge node[above] {} +(-1,0);
    \path[->] (v_r) edge node[above] {} +(1,0);
    \path[<-] (v_l) edge [bend left] node[above] {} +(2,0);
    \path[<-] (v_r) edge [bend left] node[above] {} +(-2,0);
    \path[<->] (v_r) edge node[above] {$c_{\text{rev}}$} (v_l);
    
\end{tikzpicture}
\caption{One node $v$ on the left is split into two nodes $v_l$ and $v_r$ on the right. This is done in order to keep track of the direction of movement for agents. Agents that would move onto $v$ from the right side will arrive on $v_r$, indicating rightward movement. Should an agent reverse direction, a cost of $c_{\text{rev}}$ can be applied.}
\label{fig:dirofmov}
\end{figure}

\co{graaf model waarbij een node gesplits wordt in twee nodes om extra kosten toe te wijzen aan omkeren}
We propose to model this restriction of a train's movement by adapting the graph. Every node where a train is allowed to reverse its direction is split in two nodes as illustrated in Figure~\ref{fig:dirofmov}. Nodes $v$ on which a train is not allowed to reverse direction are split in the same manner, except there is no edge between $v_l$ and $v_r$. For every pair of nodes $v_l$ and $v_r$ it is required that only one agent can occupy either node.

\subsection{Including service tasks}
A further important extension of MAPF to model TUSS more accurately is the inclusion of service tasks. This is the most prevalent reason that trains are required to not only park, but also move over the yard throughout the maintenance period. This extension is thus aimed at improving the capacity bound.

During the maintenance period trains are checked, cleaned, repaired, and parked at the shunting yard. These tasks will be referred to as service tasks. Each train has a set of service tasks that it has to complete during the maintenance period. Each service task can be performed at several locations.

\co{TUSS has multiple maintenance tasks for each train, Tasks can be performed at multiple locations, tasks have pre-specified durations, the order of tasks is not specified,  proposed model}
We propose to model the service tasks in the form of waypoints that agents should pass at some moment during their path. 
Each agent $a\in A$ is assigned a set of service tasks $S_a=\{s_1,\hdots,s_k\}$, where each service task $s_i$, for $i\in \{1,\hdots,k\}$, corresponds to a set of locations $L_{i}\subset L$ at which the service task can be completed. 
%For each such task we also have a duration $\tau_i$. 
An agent that is being serviced has to remain at the service location for the duration $\tau_i$ of the service task. Therefore, a service task $s_i\in S_a, i\in \{1,\hdots,k\}$ is said to be completed when the path $p\in P_a$ of agent $a$ includes the same location $l\in L_i$ for $\tau_i$ consecutive time steps. A path $p\in P_a$ is only considered feasible when all maintenance tasks of agent $a$ are completed. By defining the service tasks in this manner we ensure two properties: 1) the order of the service tasks is not of pre-defined, and 2) agents can move over unoccupied nodes that are also service locations just like any other nodes.

% praktijk verbinden aan restrictie naar 1 waypoint of 1 locatie
% more significant reduction of complexity is obtained when the order of maintenance tasks is fixed. people are used to a certain order, but this is not a strict relaxation of the problem any longer.

\subsection{Other extensions and open questions}\label{sec:otherextensions}
Next we discuss several more practical extensions we believe to be of less importance.
%, for which we do not propose details on how to model them. 

\co{parking only on certain tracks}
The first of such extensions is that some locations in the shunting yard are not eligible for parking, such as track changes and platforms. Trains can only remain on such locations for several time steps at most. 
%On several parts of the shunting yard, on track changes for example, train units are not allowed to park for a prolonged period of time. Trains can only stay there for several time steps for example. 
A strict way in which this can be modeled is to remove all wait edges for the corresponding vertices. This might however be too strict, such that the model is no longer a relaxation of the real-life TUSS application.
%\co{abstract model for long tracks (corridors) -- (when) is this necessary, what are straightforward alternatives (e.g. discretisation such as in grids) and their (dis)advantages (first assume single-units only)} does this fit in this paper? I'm not sure.

\co{couple/decouple train units from/into composite trains}
For the moment we also ignore the composition of train units into trains and the related coupling and decoupling activities.
%In practice trains are moved in compositions. These compositions can be split and joined. 
Train units (just as agents) are atomic. Maintenance tasks are assigned to train units. 
%Trains are persistent. 
Trains are not atomic. Arriving events and departure goals are defined for trains.
%A required train (composition) may not yet exist (and if matching is part of the problem, then a matching outgoing train combination needs to be found). 
It is still an open question what an effective extension of MAPF is to represent this.
Without this, however, the current model is a relaxation.

\co{different train lengths -- so parking tracks cannot be modeled as nodes that allow fixed number of trains}
A basic way in which tracks in the shunting yard can be represented by a graph is to divide every track into multiple nodes. A natural choice would be to divide the track length by the length of the shortest train type and round up to obtain the number of nodes that together represent that track. This method ensures there is at least as much space for trains to move around in the model as there is on the represented shunting yard, resulting in a relaxation. To obtain a more accurate representation, which is more strict, the train length can be paired with train types. This assigns a length to every train unit.
A track can be modelled by one or more nodes, where a constraint needs to ensure that the sum of the lengths of the trains on a track is not more than the length of the nodes.
Additionally, such a representation should then accurately reflect the train order on the track to guarantee that only trains at the beginning or the end of the track can leave the track, and also only in that direction.
%This can for example be done using a double-ended queue.
%New trains cannot enter the track if there is no place left on the track to park. To model this, one needs to choose how realistic the modelling of the parking position on the track should be.
    
\co{allocation of personnel (including walking times, constraints on working/driving times)}
An extension that is very much part of the real-life problem is the planning of personnel. Personnel is required in the form of train drivers, maintenance crews and cleaners. 
Employees of each category are limited in their number, ability and working hours.
Operational costs are very high if all trains have their own driver for the duration of the entire maintenance period and the maintenance period extends for longer than one shift. Also, employees can only perform certain tasks. In all likelihood, the train drivers do not have the know-how required for a technical maintenance check of the train, the maintenance engineers are unaware of how to properly clean the train, and the cleaners cannot drive the train around the shunting yard. Moreover, when crew has to move from one service track to another, or from one train to another, walking times have to be taken into account.
If indeed there are known limitations on the available personnel, then scheduling personnel is a very important extension.
However, when the main objective of the model is to provide a bound on the capacity of the (very expensive) yard infrastructure, this extension is not necessarily a relaxation.

\co{more detailed timing; e.g., braking, speeding, maximum speeds, safety constraints on crossings}
Lastly, more detail could be added in terms of physical correctness. The speed of trains can vary during movement. Because trains have such a high mass, changing the momentum by accelerating or breaking can take a significant time. There are also safety measures that limit the speed of trains and install a minimum time between two trains crossing the same track change. The way in which these extensions are implemented should conserve the relaxation property of the model.

\section{Conclusion and Future work} %beter kopje?
\label{sec:conclusions}
\co{intro}
%In this final section, we provide a discussion of the proposed extensions, we conclude our analysis of the problem and its extensions, and propose directions for future work.

%\co{does an optimal solution provide a bound on the real world feasibility? (e.g. is the model itself a relaxation?)}
The real-life application of the Train Unit Shunting and Servicing problem contains many practical details that are hard to account for in an abstract model. We aim to find an upper bound on the servicing capacity of a shunting yard and it is therefore important to extend the model in a manner that such a bound can be guaranteed. 
%The matching extension is crucial in this regard, for without it, the model is not a relaxation.

\co{conclusions}
The main conclusion from the presented analysis is that it is essential to first extend MAPF with some representation of the matching problem.
This, in its simplest form has strong similarities to the task allocation and planning MAPF models discussed in related work~\citep{Ma2016,Andreychuk2019Multi-agentTime}.
If done properly, this extension provides a relaxation of TUSS.

This paper described further refinements to obtain better bounds:
time-specific arrival and departure times can be modeled such that agents no longer start moving towards their goals at the same time, but instead enter the shunting yard at the arrival time and leave shortly before their departure from one of the platforms.
Keeping track of the direction of movement of agents can be achieved by splitting each node into two nodes. Each new node then resembles one direction of movement. The capacity of the new nodes is in total still only enough to harbour one agent at a time.
The addition of service tasks in the form of waypoints resembles the need for trains to be checked, cleaned and repaired at pre-defined locations on the shunting yard. Each agent has their own set of service tasks to complete, and each service task can be completed at a given set of locations. The service tasks also take a specified amount of time to complete, in which the agent has to remain on the service location.
We provide arguments why each of the resulting variants is a relaxation of TUSS. By formulating these extensions precisely, we aim to interest researchers in this real-life problem and leverage existing work on MAPF to solve TUSS.

\co{other extensions in 4.4 still require to be formalized}
Keeping the goal of relaxation in mind, the extensions proposed in Subsection~\ref{sec:otherextensions} should be further formalized. 
%As long as these extensions are only described as concepts and suggestions, multiple variants might emerge that differ from the real-life application.
Apart from these extensions, we propose several other interesting directions for future research.

\co{use to find feasible solutions -- so restrictions instead of relaxations}
A similar exercise as reported in this paper could be done for the use of finding feasible solutions to the TUSS problem. This would require the model variants to be restrictions rather than relaxations. The big advantage is that any feasible solution to a restriction can be translated into a feasible solution of the original problem, so a complete or optimal algorithm is not necessary.

\co{introducing disturbances}

Another future research direction is to introduce disturbances into the problem. 
%The earlier mentioned Flatland challenge competition \citep{flatland2020} considered disturbances (temporary train break-downs) in a later round. However, this competition does not include the matching problem or service tasks. 
For the TUSS the exploration of so-called robust schedules seeks to construct the maintenance schedule such that disturbances in arrival times, train or track defects and personnel running late, can be absorbed by making no or only a very small change to the schedule.
Although, there are no papers that provide robust schedules for the TUSS problem, 
several robustness measures to measure the robustness of TUSS schedules are proposed by \cite{VanDenBroek2018HowPlans}.  

\co{objective: is makespan or sum of path costs the most relevant objective?  number of train drivers required seems a better one, but these are not in our model (currently), so maybe: number of simultaneous movements, or number of ``startups``}
Finally, we have considered MAPF and its extensions mainly as a feasibility problem. In practice, costs emerge mostly from the personnel that is required to fulfill all the scheduled tasks. It would therefore be beneficial to minimize the number of simultaneous actions, for example with regards to train movements.

\bibliographystyle{apalike}
\bibliography{TUSP-Mendeley,TUSSP}

\end{document}